\documentclass[a4paper,11pt]{article}
\usepackage{jheppub}
\makeatletter\gdef\@fpheader{}\makeatother
\usepackage[T1]{fontenc}
\usepackage{enumitem}
\usepackage{orcidlink}
\usepackage{hepparticles}
\usepackage{xspace}
\usepackage{amsmath}

\newcommand{\ie}{\mbox{i.e.}\xspace}
\newcommand{\eg}{\mbox{e.g.}\xspace}

\newcommand{\unit}[1]{\ensuremath{\text{\,#1}}\xspace}
\newcommand{\GeV}{\ensuremath{\,\text{Ge\hspace{-.08em}V}}\xspace}
\newcommand{\TeV}{\ensuremath{\,\text{Te\hspace{-.08em}V}}\xspace}
\newcommand{\fbinv}{\ensuremath{\,\text{fb}^{-1}}\xspace}

\newcommand{\abs}[1]{\ensuremath{\lvert #1 \rvert}}
\newcommand{\abseta}{\ensuremath{\abs{\eta}}\xspace}

\newcommand{\DR}{\ensuremath{\Delta R}\xspace}
\newcommand{\HT}{\ensuremath{H_{\mathrm{T}}}\xspace}

\newcommand{\pt}{\ensuremath{p_{\mathrm{T}}}\xspace}

\newcommand{\PQb}{{\HepParticle{b}{}{}}\xspace}
\newcommand{\PQt}{{\HepParticle{t}{}{}}\xspace}
\newcommand{\PAQb}{{\HepAntiParticle{\PQb}{}{}}\xspace}
\newcommand{\PAQt}{{\HepAntiParticle{\PQt}{}{}}\xspace}
\newcommand{\PW}{{\HepParticle{W}{}{}}\xspace}
\newcommand{\PWp}{{\HepParticle{\PW}{}{+}}\xspace}
\newcommand{\PWm}{{\HepParticle{\PW}{}{-}}\xspace}
\newcommand{\etat}{{\HepParticle{\eta}{\PQt}{}}\xspace}
\newcommand{\ttbar}{\ensuremath{\PQt\PAQt}\xspace}

\newcommand{\obs}{\ensuremath{\mathcal{O}}\xspace}
\newcommand{\dSdO}{\ensuremath{\frac{\mathrm{d}\sigma}{\mathrm{d}\obs}}\xspace}
\newcommand{\dSdOinl}{\ensuremath{\mathrm{d}\sigma/\mathrm{d}\obs}\xspace}
\newcommand{\xO}{\ensuremath{x_{\obs}}\xspace}
\newcommand{\xOtilde}{\ensuremath{\tilde{x}_{\obs}}\xspace}
\newcommand{\xOvar}[1]{\ensuremath{x_{\obs\!,#1}}\xspace}
\newcommand{\poi}{\ensuremath{c}\xspace}

\newcommand{\DELPHES}{{\textsc{delphes}}\xspace}

\newcommand{\MADSPIN}{{\textsc{madspin}}\xspace}
\newcommand{\MGvATNLO}{{\textsc{MadGraph}5\_a\textsc{mc@nlo}}\xspace}

\newcommand{\PYTHIA}{{\textsc{pythia}}\xspace}
\newcommand{\TOPpp}{{\textsc{top++}}\xspace}

\newcommand{\mtt}{\ensuremath{m_{\ttbar}}\xspace}
\newcommand{\chel}{\ensuremath{c_{\mathrm{hel}}}\xspace}
\newcommand{\chan}{\ensuremath{c_{\mathrm{han}}}\xspace}

\title{An Optimal Observable Machine for reinterpretable measurements in high-energy physics}

\author[a,b]{Torben Mohr\,\orcidlink{0009-0006-8863-8683},}
\author[a]{Alejandro Quiroga Trivi\~no\,\orcidlink{0000-0001-8570-1970},}
\author[a]{Fabian Riemer\,\orcidlink{0009-0004-1975-3851},}
\author[a]{Artur Monsch\,\orcidlink{0009-0007-3529-1644},}
\author[c]{Matteo Defranchis\,\orcidlink{0000-0001-9573-3714},}
\author[a]{Joscha Knolle\,\orcidlink{0000-0002-4781-5704},}
\author[c]{Ankita Mehta\,\orcidlink{0000-0002-0433-4484},}
\author[a]{Jan Kieseler\,\orcidlink{0000-0003-1644-7678},}
\author[a]{and Markus Klute\,\orcidlink{0000-0002-0869-5631}}

\affiliation[a]{%
    Karlsruhe Institute of Technology (KIT), Karlsruhe, Germany
}
\affiliation[b]{%
    Ghent University, Ghent, Belgium
}
\affiliation[c]{%
    European Organization for Nuclear Research (CERN), Geneva, Switzerland
}

\emailAdd{torben.mohr@cern.ch}
\emailAdd{alejandro.quiroga.trivino@cern.ch}
\emailAdd{fabian.riemer@student.kit.edu}
\emailAdd{artur.artemij.monsch@cern.ch}
\emailAdd{matteo.defranchis@cern.ch}
\emailAdd{joscha.knolle@cern.ch}
\emailAdd{ankita.mehta@cern.ch}
\emailAdd{jan.kieseler@cern.ch}
\emailAdd{markus.klute@cern.ch}

\abstract{A machine-learning-based framework for constructing generator-level observables optimized for parameter extraction in particle physics analyses is introduced, referred to as the Optimal Observable Machine (OOM).
Unfoldable differential distributions are learned that maximize sensitivity to a parameter of interest while remaining robust against detector effects, systematic uncertainties, and biases introduced by the unfolding procedure.
Detector response and systematic uncertainties are explicitly incorporated into the training through a likelihood-based loss function, enabling a direct optimization of the expected measurement precision while minimizing the bias from any assumption on the parameter of interest itself.
The approach is demonstrated in an application to top quark physics, focusing on the measurement of a recently observed pseudoscalar excess at the top quark pair production threshold in dilepton final states.
It is shown that a generator-level observable with enhanced sensitivity and long-term reinterpretability can be constructed using this method.}

\begin{document}
\maketitle
\flushbottom

\section{Introduction}

Extracting physical parameters from experimental data is a central task in particle physics.
This includes precision measurements of standard model (SM) parameters and searches for new physics, which are often interpreted as constraints on beyond-the-SM models.
In both cases, parameters are inferred by comparing the measured distributions of carefully chosen observables with theoretical predictions, taking into account both the sensitivity of the observable and the achievable experimental precision.

Traditionally, observables are constructed based on physical intuition or analytical considerations and unfolded to the parton or particle level (\ie, defined purely in terms of parton or particle information before detector reconstruction, also referred to as ``generator level''), enabling direct comparison with theoretical predictions.
Alternatively, templates can be generated at the level of reconstructed objects using detector simulation (``detector-level''), often offering higher sensitivity~\cite{CMS:2024irj}.
However, performing the detector simulation is computationally expensive and at high precision generally only feasible within experimental collaborations~\cite{GEANT4:2002zbu, ATLAS:2010arf, CMS:NOTE-2022-008}.
As a result, reinterpretations by the broader community typically rely on unfolded measurements, \eg, Refs.~\cite{Celada:2024mcf, deBlas:2025xhe} for recent examples.

The increasing use of machine-learning methods has made detector-level analyses more powerful, but also more complex~\cite{Schwartz:2021ftp}.
While such approaches have enabled increasingly precise parameter extractions, their complexity often prevents straightforward reinterpretation, limiting the long-term usability, or ``shelf life'', of the resulting measurements.

{\tolerance=800
In this work, we introduce the \emph{Optimal Observable Machine} (OOM), a machine-learning-based framework for constructing unfoldable observables that are optimally suited for parameter extraction.
Rather than selecting observables \textit{a priori}, the OOM \textit{learns} an unfoldable differential distribution that is explicitly tailored to a given parameter of interest.
The learned observable is designed to maximize sensitivity to the parameter while remaining robust against detector effects, systematic uncertainties, and biases introduced by the unfolding procedure.
As a result, the OOM yields unfolded generator-level measurements that combine high precision with long-term reinterpretability.
\par}

Top quark physics plays a central role in the study of fundamental interactions.
As the heaviest known elementary particle, the top quark is a sensitive probe of important SM parameters, such as the top quark mass, as well as of potential new physics at higher energy scales~\cite{FerreiradaSilva:2023mhf}.
To illustrate the capabilities of the OOM, we apply it to a recent example of top quark analyses performed by the ATLAS and CMS Collaborations.
Using proton-proton collision data recorded at $\sqrt{s}=13\TeV$, the experiments reported the observation of a pseudoscalar excess at the top quark pair (\ttbar) production threshold~\cite{CMS:2025kzt, CMS:2025dzq, ATLAS:2025mvr}, which can be interpreted as a ``toponium'' bound state and/or attributed to a new pseudoscalar boson~\cite{Fuks:2024yjj, Garzelli:2024uhe, Maltoni:2024tul, Bahl:2025you, Nason:2025hix, Matsuoka:2025jgm, Fuks:2025toq, Flacke:2025dwk, Arcadi:2025grl}.
We employ the OOM to obtain a generator-level observable with optimal sensitivity to distinguish between the pseudoscalar excess and the continuum \ttbar production contribution.

The use of machine-learning methods to construct optimal detector-level observables is a well-established technique in measurements and searches of the ATLAS and CMS experiments~\cite{ATLAS:2024lda, ATLAS:2024fdw, CMS:2024phk, ATLAS:2024fkg, ATLAS:2024wla, ATLAS:2024kxj, ATLAS:2024itc, CMS:2024krd, CMS:2024zqs, CMS:2024bni, CMS:2024gzs}.
Recent machine-learning developments have incorporated systematic uncertainties in the observable optimization~\cite{Neal:2007zz, Cranmer:2015bka, DeCastro:2018psv, Wunsch:2020iuh, Simpson:2022suz, CMS:2025cwy}.
We build the OOM on top of the approach from Ref.~\cite{CMS:2025cwy}, which we extend to the optimization of generator-level observables.

\section{Observable finding}
\label{sec:method}

The goal of the OOM is to find an observable \obs with a generator-level differential cross section \dSdOinl that simultaneously fulfills the following goals:
\begin{enumerate}[label=(\roman*)]
\item\label{item:genlevel} It is defined exclusively through generator-level information.
\item\label{item:poisensitive} It maximizes sensitivity to the parameter of interest \poi.
\item\label{item:measureable} It is precisely measurable in the presence of systematic uncertainties.
\item\label{item:modelindependent} It has minimal dependence on theoretical uncertainties and model assumptions.
\end{enumerate}
The corresponding detector-level distribution is denoted by \xO, such that \dSdOinl can be measured by applying unfolding to \xO~\cite{Cowan:2002in}.
In an idealized scenario without background contributions, the response matrix $R$ encodes all detector and reconstruction effects relevant for the unfolding, such that the following relation holds:
\begin{equation}\label{eq:unfolding}
    \xO(\poi,\omega)=R(\poi,\omega)\,\dSdO(\poi,\omega).
\end{equation}
Here, \dSdOinl and \xO are to be understood as vectors with entries representing the bins of the two distributions.
Generally, all three quantities \dSdOinl, \xO, and $R$ depend on \poi, as well as on nuisance parameters representing the effect of various sources of systematic uncertainties, denoted collectively by $\omega$.
The number of generator- and detector-level bins is not required to be the same, and thus $R$ is not necessarily square.

Instead of constructing the observable by hand, we determine \dSdOinl through a learnable deterministic function $\Phi_g$ as:
\begin{equation}
    \dSdO(c,\omega)=\mathcal{H}\big[\Phi_g\big(f_g(\poi,\omega)\big)\big],
\end{equation}
where $\mathcal{H}$ denotes the histogramming (or binning) operation and $f_g$ represents a set of generator-level features.
These features have to be selected such that they provide sensitivity to \poi, which is how the sensitivity of \dSdOinl to \poi is achieved.
However, the features generally also depend on the systematic uncertainties $\omega$, and thus cause a dependence of \dSdOinl on $\omega$.
Through this construction, goal~\ref{item:genlevel} is automatically fulfilled, and any reinterpretation of a measurement of \dSdOinl can be performed without requiring detector simulation.
Similarly, we determine \xO as:
\begin{equation}
    \xO(c,\omega)=\mathcal{H}\big[\Phi_d\big(f_d(\poi,\omega)\big)\big],
\end{equation}
depending on a set $f_d$ of detector-level features, with similar considerations for the dependence on \poi and $\omega$ as for $f_g$.
In practice, $\Phi_g$ and $\Phi_d$ are implemented as multilayer perceptrons (MLPs) and trained simultaneously.

To achieve goal~\ref{item:poisensitive}, we consider a loss function for the MLP training that rewards a precise extraction of \poi from the optimized observable.
Starting from the binned likelihood function $\mathcal{L}$ that represents the measurement of \poi, we calculate the Hessian matrix
\begin{equation}
    H_{\alpha\beta}=\frac{\partial^2}{\partial\alpha\,\partial\beta}\big({-}\log\mathcal{L}\big),
\end{equation}
where the derivatives are taken with respect to \poi and the nuisance parameters, \ie, $\alpha,\beta\in\{\poi,\omega\}$.
We can then calculate the expected uncertainty in \poi, denoted by $\Delta\poi$, from the diagonal element of the inverted Hessian matrix:
\begin{equation}\label{eq:loss}
    \Delta\poi=\sqrt{H_{\poi\poi}^{-1}},
\end{equation}
which is used directly as the loss function during the training.
This approach, described in more detail in Ref.~\cite{CMS:2025cwy}, utilizes the Fisher information formalism~\cite{Fisher:1922saa} for the case of the Asimov data set~\cite{Cowan:2010js}, \ie, when considering the expected sensitivity.

The likelihood function for a binned profile likelihood fit with nuisance parameters is formulated as:
\begin{equation}\label{eq:recolikelihood}
    \mathcal{L}=\mathcal{L}_p\big(d\,\big\vert\,\xO(\poi,\omega)\big)\cdot\mathcal{L}_{\mathrm{NP}}(\omega),
\end{equation}
where $\mathcal{L}_p$ represents the Poisson likelihood to obtain the observed data distribution, denoted by $d$, given the prediction $\xO(\poi,\omega)$, and $\mathcal{L}_{\mathrm{NP}}$ encompasses penalty terms for all nuisance parameters $\omega$~\cite{ATLAS:2011tau, CMS:2024onh}.
Incorporating systematic uncertainties through $\mathcal{L}_{\mathrm{NP}}$ in the loss function is a central aspect of the \emph{systematic-uncertainty-aware training} as introduced in Ref.~\cite{CMS:2025cwy}, ensuring that the training goal aligns with the final precision of the measurement of \poi even in cases where systematic uncertainties become comparable to or dominant over statistical uncertainties.

Training $\Phi_d$ with a loss function based on Eq.~\eqref{eq:recolikelihood} would yield a detector-level distribution with maximal sensitivity to \poi that is precisely measurable in the presence of systematic uncertainties.
We extend this approach to an \emph{unfolding-aware training} by substituting \xO via Eq.~\eqref{eq:unfolding} in the likelihood function:
\begin{equation}\label{eq:mainlikelihood}
    \mathcal{L}=\mathcal{L}_p\bigg(d\,\bigg\vert\,R(\poi,\omega)\,\dSdO(\poi,\omega)\bigg)\cdot\mathcal{L}_{\mathrm{NP}}(\omega),
\end{equation}
This way, we phrase the training as a search for the generator-level distribution \dSdOinl that has maximum sensitivity to \poi~\ref{item:poisensitive} in the presence of systematic uncertainties~\ref{item:measureable}, including those that affect the generator-level distribution~\ref{item:modelindependent}.
Since $R$ is determined from the definitions of \dSdOinl and \xO, the resulting loss function is sensitive to the definition of \xO as well.

{\tolerance=800
A critical aspect of the method is to ensure gradient flow through all learnable parameters.
We aim for a single-utility optimization based on Eq.~\eqref{eq:mainlikelihood}.
At each gradient descent step, $\Phi_g$ and $\Phi_d$ are evaluated and the response matrix is determined from their binned output.
We then calculate:
\begin{equation}
    \xOtilde(\poi,\omega)=\frac{1}{2}\Big\{R(\poi,\omega)\,\mathcal{H}\big[\Phi_g\big(f_g(\poi,\omega)\big)\big]+\mathcal{H}\big[\Phi_d\big(f_d(\poi,\omega)\big)\big]\Big\},
\end{equation}}
and evaluate
\begin{equation}\label{eq:finallikelihood}
    \mathcal{L}=\mathcal{L}_p\big(d\,\big\vert\,\xOtilde(\poi,\omega)\big)\cdot\mathcal{L}_{\mathrm{NP}}(\omega).
\end{equation}
While the result is identical compared to the usage of Eq.~\eqref{eq:mainlikelihood} directly with respect to the forward evaluation, this procedure introduces explicit gradients into both $\Phi_g$ and $\Phi_d$ from the training on detector level.

A technical aspect that requires special care in this context is the treatment of histogram binning.
Since the loss function is defined in terms of binned distributions, gradients must be propagated through the binning operation in order to enable end-to-end training.
This necessitates a differentiable formulation of the binning step, which assigns events to bins in a smooth manner while preserving the interpretation of the resulting histograms as approximate counts.
We use the approach proposed in Ref.~\cite{CMS:2025cwy} for the one-dimensional distributions as well as for the two-dimensional response matrix.

In any unfolding approach, prior assumptions on \poi have to be made in order to calculate $R$.
Since $R$ generally carries a dependence on \poi, this can lead to a bias in the unfolded distribution.
With $\poi_0$ being the nominal hypothesis and $\poi_{\alpha}$ representing a set of alternative values, we can evaluate the folded distributions $\xOvar{\alpha}=R(\poi_{\alpha})\,\dSdOinl(\poi_0)$, \ie, folding the nominal generator-level distribution with the different hypotheses for the response matrices, and quantify the bias by calculating the difference between \xOvar{\alpha} and the nominal result \xOvar{0}.
To minimize the dependence of $R$ on \poi and ensure a good reinterpretability of the unfolded we result, we add the following penalty term to the loss function used in the OOM:
\begin{equation}\label{eq:responseconstraint}
    \lambda\exp\sum_{\alpha}\chi^2(\xOvar{\alpha},\xOvar{0}),
\end{equation}
where $\lambda$ is a parameter that governs the strength of the constraint.
This way, the training favors observables for which $R$ is approximately independent of \poi, meaning that the sensitivity on \xO on \poi must predominantly originate from \dSdOinl through Eq.~\eqref{eq:unfolding}.

Finally, background contributions can be incorporated in a straightforward manner.
Backgrounds that are independent of \poi are added at the detector level to \xO, including their dependence on $\omega$.
If a background contribution itself depends on \poi, it may instead be treated as part of the signal, either by incorporating it into the learned generator-level distribution or by modeling it through an additional response matrix.
It is also possible to introduce regularization into the unfolding procedure if desired in the usual way by adding additional penalty terms~\cite{Cowan:2002in}.

\section{Optimal observable for toponium}

We apply the OOM to the measurement of toponium in dilepton final states.
In Refs.~\cite{CMS:2025kzt, CMS:2025dzq, ATLAS:2025mvr}, the ATLAS and CMS Collaborations construct three-dimensional distributions at detector level using:
\begin{itemize}
    \item the reconstructed invariant mass of the top quark pair, denoted by \mtt;
    \item the scalar product of the unit vectors of the momenta of the two leptons in the rest frames of their parent top quark and antiquark, respectively, denoted by \chel;
    \item a similar scalar product but with the sign of the component parallel to the top quark direction flipped for one of the leptons, denoted by \chan.
\end{itemize}
The spin correlation variables \chel and \chan provide sensitivity to distinguish between a scalar or pseudoscalar contribution of the observed excess, whereas \mtt provides sensitivity to the mass of the excess.
We consider the signal strength $r$ of the toponium sample, \ie, the ratio of the measured cross section to the prediction, as the parameter of interest (\ie, \poi in the notation of Section~\ref{sec:method}), and demonstrate how to find an optimal observable both at generator- and detector-level to constrain $r$ in the presence of systematic uncertainties.

\subsection{Simulated event samples}

The OOM is applied using simulated event samples including a fast detector simulation.
We employ \MGvATNLO2.6.5~\cite{Alwall:2014hca} for the matrix-element calculation, the NNPDF3.1NLO~\cite{NNPDF:2017mvq} parton distribution functions, \PYTHIA8.310~\cite{Bierlich:2022pfr} for the simulation of parton shower, fragmentation, and hadronization, and \DELPHES3.5.0~\cite{deFavereau:2013fsa, Selvaggi:2014mya, Mertens:2015kba} with its default CMS detector configuration for the detector response and pileup effects.
The considered scenario is the \mbox{Run 3} data-taking period of the LHC, \ie, we simulate proton-proton collisions at $\sqrt{s}=13.6\TeV$ and assume an integrated luminosity of 300\fbinv for the normalization of the event counts.

Samples for \ttbar production at next-to-leading order in perturbative quantum chromodynamics (simply referred to as ``\ttbar sample'' in the following) and for the production of a pseudoscalar resonance \etat in the final state $\etat\to\PWp\PQb\PWm\PAQb$ (``\etat sample'') are generated.
A value of 172.5\GeV is used for the top quark mass.
Top quark decays in the \ttbar samples are simulated with \MADSPIN~\cite{Artoisenet:2012st}, and the sample is normalized to a predicted cross section of $924\,^{+32}_{-40}\unit{pb}$, as calculated with \TOPpp2.0~\cite{Czakon:2011xx}.
The \etat sample uses a simplified model for a generic resonance~\cite{Fuks:2021xje, Maltoni:2024tul} with mass set to 343\GeV (twice the top quark mass minus a binding energy of about 2\GeV~\cite{Fuks:2021xje}) and width to 2.8\GeV (twice the top quark width~\cite{Fadin:1990wx}), following the setup in Ref.~\cite{CMS:2025kzt}.
It is normalized to a cross section of 7.1\unit{fb}, obtained from scaling the predicted 13\TeV cross section of 6.4\unit{fb}~\cite{Sumino:2010bv, Fuks:2021xje, CMS:2025kzt} with the ratio of the \ttbar cross sections at 13.6 and 13\TeV~\cite{Czakon:2011xx}.

At the generator level, a fiducial selection is applied requiring exactly two electrons and/or muons with opposite charge and at least two jets, of which at least one is matched to a \PQb quark.
Similarly, an event selection is applied at the detector level.
In both selections, electrons are required to have transverse momentum $\pt>25\GeV$ and pseudorapidity $\abseta<2.5$, muons to have $\pt>25\GeV$ and $\abseta<2.4$, and jets to have $\pt>30\GeV$, $\abseta<2.5$, and an angular separation $\DR=\sqrt{\smash[b]{(\Delta\eta)^2+(\Delta\phi)^2}}>0.4$ from any selected lepton, where $\Delta\eta$ and $\Delta\phi$ are the differences in $\eta$ and azimuthal angle, respectively.
To emulate an efficient lepton identification, we additionally require detector-level electrons and muons to be geometrically matched to a selected generator-level lepton.

\subsection{Training setup}

We train two MLPs $\Phi_g$ and $\Phi_d$, implemented in \textsc{PyTorch}~\cite{Paszke:2019xhz} with three hidden layers with 100 nodes each and a ReLU activation function~\cite{Nair:2010icml} after each layer.
The output layer uses a sigmoid activation function.
Following Ref.~\cite{CMS:2025cwy}, we first perform a pretraining of each MLP to optimize the separation between fiducial \etat and fiducial \ttbar events.
We then perform the simultaneous training as outlined in Section~\ref{sec:method}, using fiducial events from both samples as signal contributions and nonfiducial events as background contributions.
The parameter of interest is the toponium signal strength $r$, which is only applied to the fiducial \etat events, so we can rewrite Eq.~\eqref{eq:unfolding} as:
\begin{align}
    \xO^{\ttbar}(\omega)+\xO^{\etat}(r,\omega)=R(r,\omega)\,\Bigg[\dSdO^{\!\ttbar}(\omega)+\dSdO^{\!\etat}(r,\omega)\Bigg],
\end{align}
with a common response matrix $R$ for both processes that can generally depend on $r$.

The generator-level features $f_g$ are \pt and $\eta$ of the top quark and antiquark; $\Delta\eta$, $\Delta\phi$, and \DR between the top quark and antiquark; and \mtt, \chel, and \chan evaluated with generator-level objects.
At the detector level, the reconstruction of the four-momenta of the top quark and antiquark is nontrivial.
We apply a full kinematic reconstruction (FKR)~\cite{CMS:2015rld}, which reconstructs the top quark and antiquark four momenta by imposing constraints on the \PW boson and top quark mass and finding algebraic solutions for the neutrino momenta, as well as a loose kinematic reconstruction (LKR)~\cite{CMS:2019esx}, which reconstructs \mtt only and not the individual momenta, but does not require assumptions about the top quark mass.
Since the FKR has a limited efficiency, \ie, it fails to provide a result for a relevant fraction of the events, we also use features that do not rely on top quark reconstruction.
The detector-level features $f_d$ used in the training are then:
\begin{itemize}
    \item (without relying on top quark reconstruction) \pt and $\eta$ of the two selected leptons and the up to four highest-\pt jets; $\Delta\eta$, $\Delta\phi$, and \DR between the two leptons; \DR between the two highest-\pt \PQb jets; the invariant mass of all selected leptons and jets; the scalar sum of the \pt of all selected jets and leptons (\HT); the transverse sphericity~\cite{ALICE:2012cor} of the event;
    \item (using FKR) \pt and $\eta$ of the top quark and antiquark; $\Delta\phi$ and \DR between top quark and antiquark; \mtt, \chel, and \chan;
    \item (using LKR) \mtt.
\end{itemize}

We consider a limited set of systematic uncertainties, which are taken from the list of leading uncertainty sources in the CMS analysis~\cite{CMS:2025kzt}.
As theoretical uncertainties, we include variations of the value of the strong coupling constant in the ISR and FSR simulation.
As experimental uncertainty, we include a variation of the jet energy scale with magnitude similar to those in the ATLAS and CMS experimental analyses~\cite{CMS:2016lmd, ATLAS:2020cli}.
Finally, we consider normalization uncertainties of 1\% in the integrated luminosity and of 4\% in the \ttbar cross section.

\subsection{Results from detector-level training}

\begin{figure}[!htp]
\centering
\raisebox{-4mm}{\includegraphics[width=0.475\textwidth]{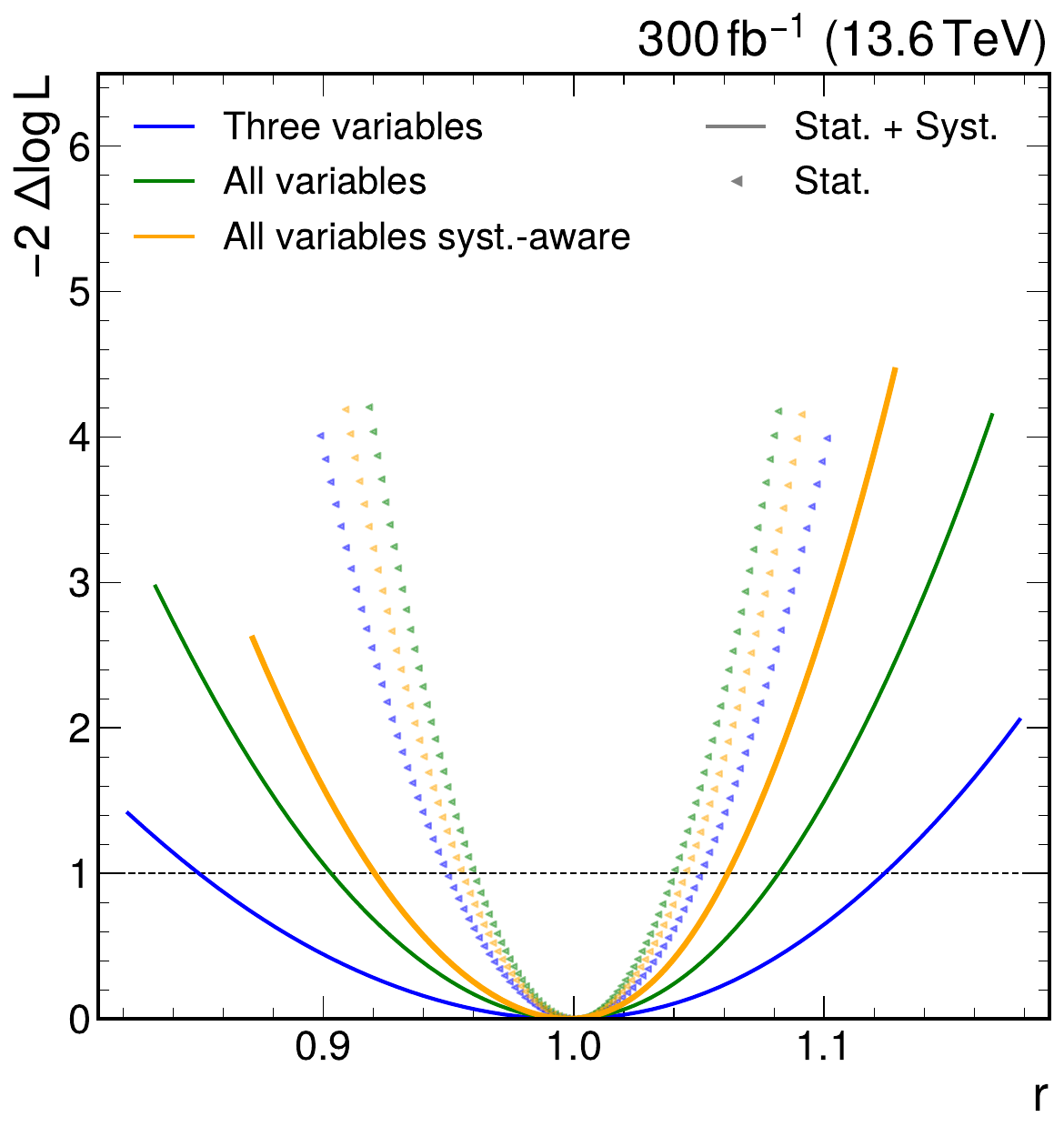}}
\hfill
\includegraphics[width=0.515\textwidth]{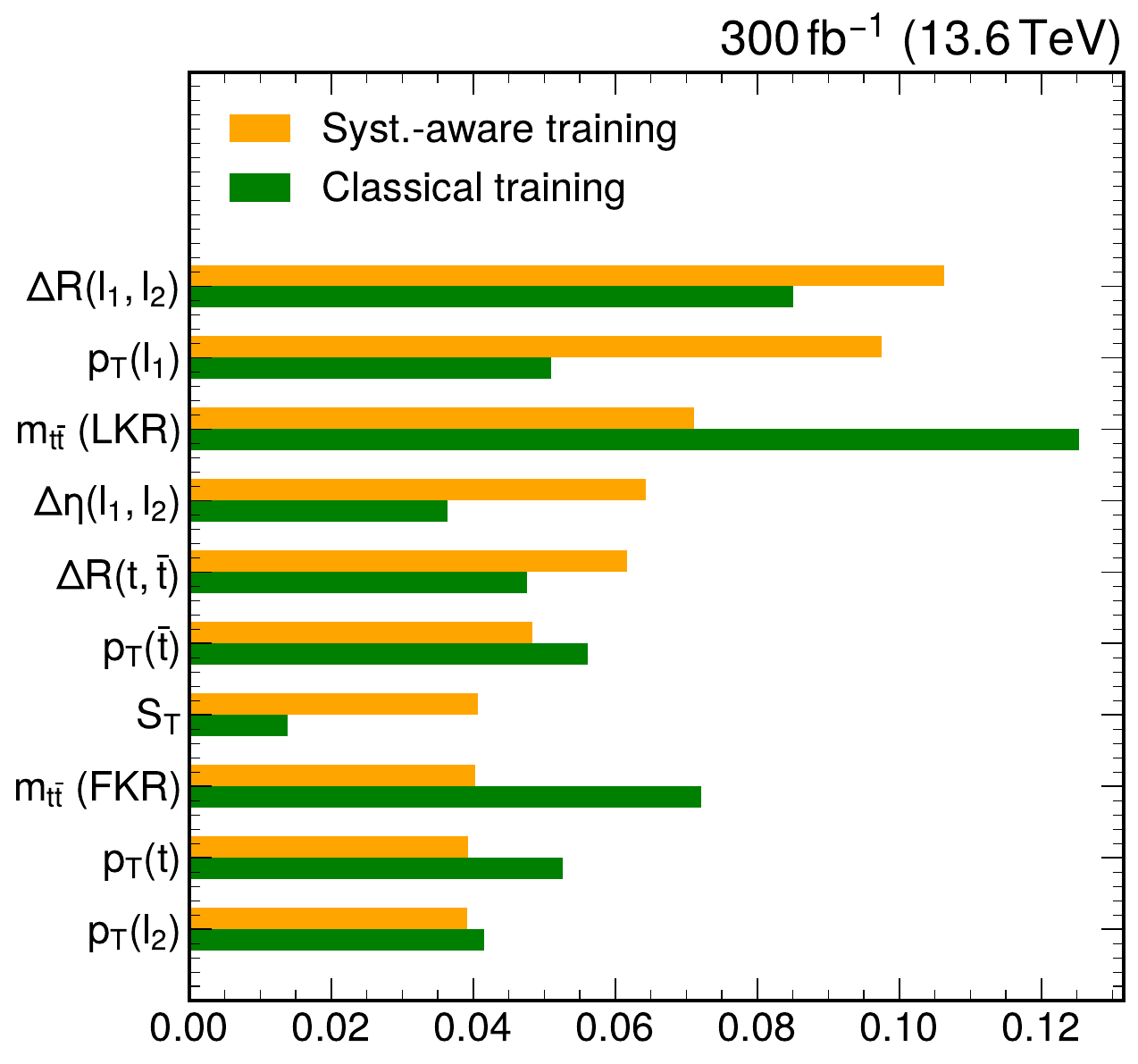}
\caption{%
    Results of different detector-level MLP training strategies.
    Left: Negative log-likelihood scans for the toponium signal strength $r$ using CE training with three input variables (blue), CE training with 29 input variables (green), and systematic-uncertainty-aware training with 29 input variables (yellow).
    Shown are the scans using only statistical uncertainties (triangles) and using both statistical and systematic uncertainties (lines).
    Right: Feature importance ranking, evaluated using Ref.~\cite{Wunsch:2018oxb}, for the ten most important input features, compared between the CE training (green) and the systematic-uncertainty-aware training (yellow), both using 29 input variables.
}
\label{fig:detectorlevelresults}
\end{figure}

We first consider only the training of $\Phi_d$ to obtain an optimal observable for the measurement of $r$ at the detector level.
As a baseline scenario, we train $\Phi_d$ with only three input variables \mtt, \chel, and \chan (evaluated using FKR), \ie, using the variables used by the original analyses~\cite{CMS:2025kzt, CMS:2025dzq, ATLAS:2025mvr}, and using cross-entropy (CE)~\cite{Good:1952rss} as loss function, which optimizes for the separation between \ttbar and \etat production considering statistical uncertainties only.
The first extended training considers the 29 input features listed above but still with CE as loss function.
For the second extension, we apply the systematic-uncertainty-aware training~\cite{CMS:2025cwy}: the loss function is defined using a likelihood function corresponding to Eq.~\eqref{eq:recolikelihood} that includes systematic uncertainty sources and is evaluated using the prediction of \xO under the nominal hypothesis for $r$ as data $d$, \ie, using the Asimov data set~\cite{Cowan:2010js}.
Results of the three training strategies are compared in Fig.~\ref{fig:detectorlevelresults}.

The negative log-likelihood scans (Fig.~\ref{fig:detectorlevelresults} left) show the precision in $r$ achieved in the different scenarios.
Using more event information significantly improves the sensitivity.
It can also be seen that the analysis, in any scenario, is limited by systematic uncertainties.
Using the systematic-uncertainty-aware training clearly improves the overall precision, even if the statistical uncertainty slightly increases.
The difference between the CE and systematic-uncertainty-aware training also become visible in the feature importance ranking (Fig.~\ref{fig:detectorlevelresults} right): while the CE training relies the strongest on \mtt evaluated using LKR, the systematic-uncertainty-aware training instead relies more heavily on the \DR between the two leptons and the highest lepton \pt, which are much less affected by the jet energy scale systematic uncertainty.

\subsection{Results from generator- and detector-level training}

\begin{figure}[!htp]
\centering
\includegraphics[width=0.55\linewidth]{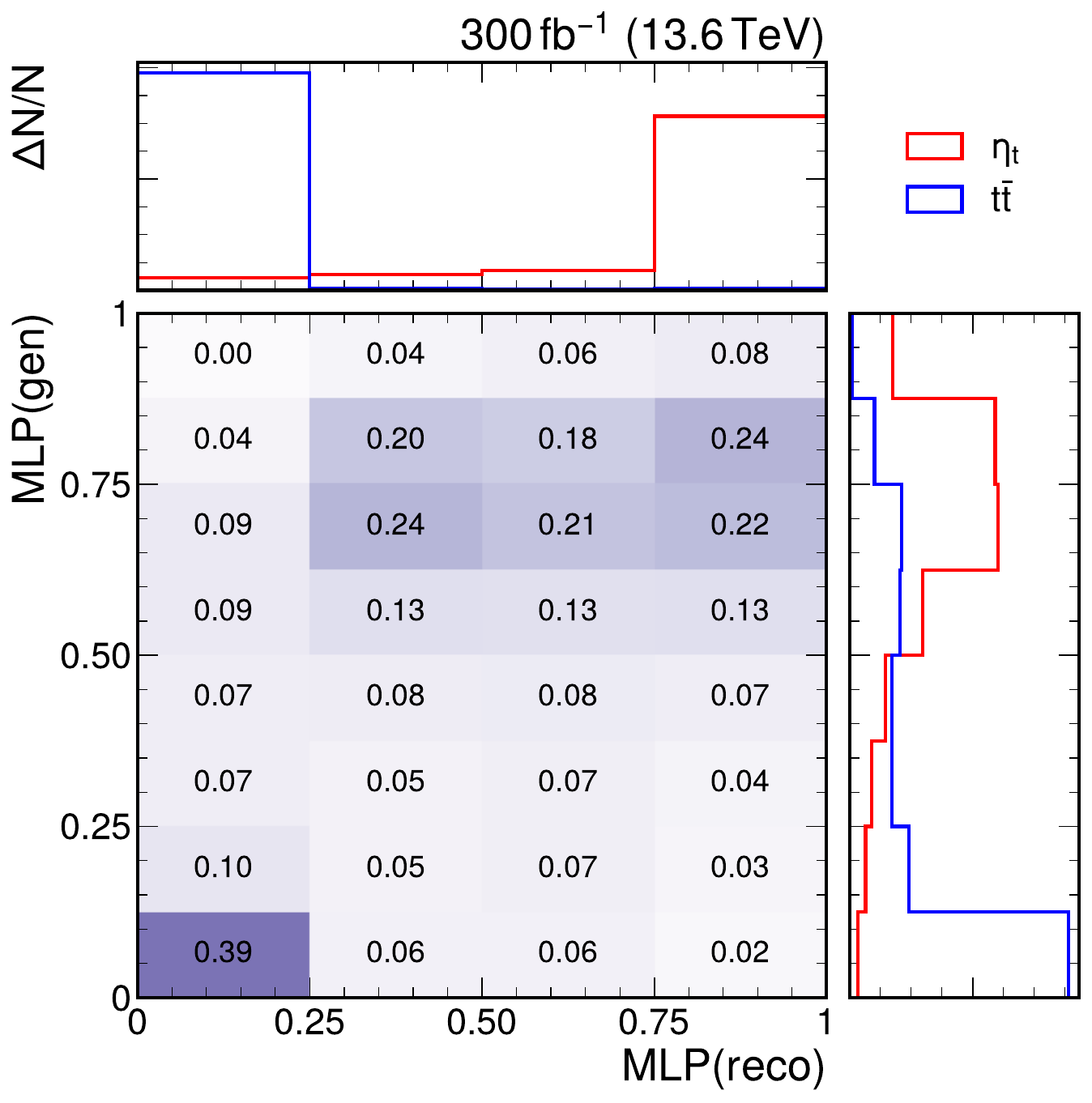}
\caption{%
    Response matrix $R$ (main panel, with generator level on the $x$ axis and detector level on the $y$ axis), normalized distribution of \dSdOinl at generator level (upper panel), and normalized distribution of \xO at detector level (right panel) for the OOM training without response-matrix constraint.
}
\label{fig:resultnoconstraint}
\end{figure}

{\tolerance=800
Now we apply the full OOM with simultaneous optimization of both generator- and detector-level distributions.
The likelihood function in Eq.~\eqref{eq:finallikelihood} is used to define the loss function.
Without the additional response-matrix constraint from Eq.~\eqref{eq:responseconstraint}, the MLPs learn distributions such that both \xO and the folded distribution $R\,\dSdOinl$ are maximally sensitive to~$r$.
The result of this training is shown in Fig.~\ref{fig:resultnoconstraint}.
\par}

\begin{figure}[!htp]
\centering
\includegraphics[width=\textwidth]{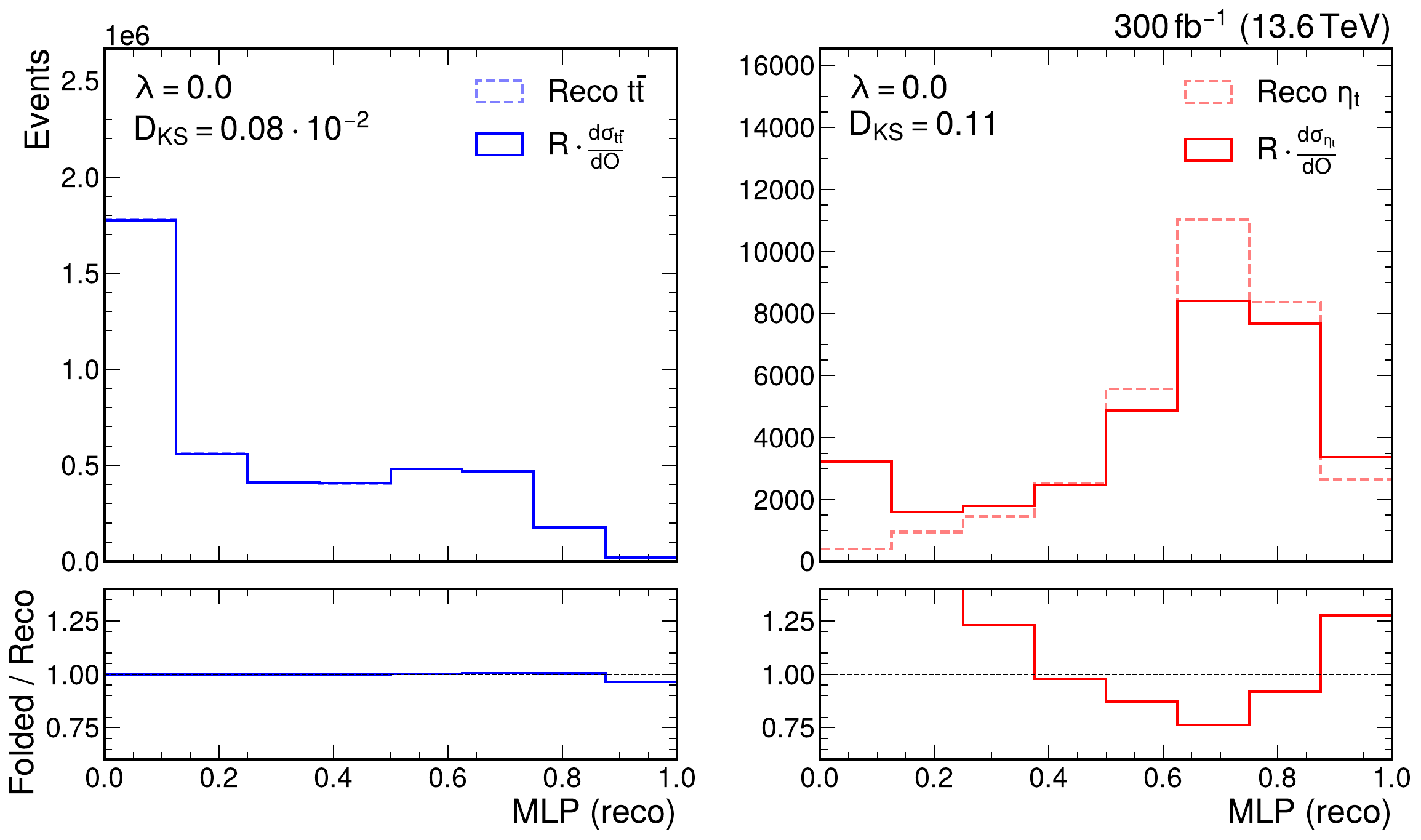}
\caption{%
    Comparison of the \ttbar (left, blue) and \etat (right, red) yields between \xO (dashed) and the folded distribution $R\,\dSdOinl$ (solid).
    The lower panel displays the ratio between the two distributions.
    Shown are the results for the training without response-matrix constraint, \ie, with $\lambda=0$.
    The captions list the Kolmogorov--Smirnov distance values.
}
\label{fig:biaslambda0p0}
\end{figure}

To evaluate a possible bias from a direct dependence of $R$ on $r$, we compare the detector-level distribution \xO with the folded distribution $R\,\dSdOinl$, separately for each of the two processes.
The result of this comparison is shown in Fig.~\ref{fig:biaslambda0p0}, providing also the Kolmogorov--Smirnov distance $D_{\mathrm{KS}}$ that quantifies the agreement between the two compared distributions.
While the bias is small for \ttbar production, which is the dominant contribution to $R$, we find a significant bias for \etat production.
Since the total yield of selected \ttbar events is more than 100 times larger than that of selected \etat events, it is clear that the response matrix is mainly derived from the \ttbar events, which explains the bias found for the \etat sample.

\begin{figure}[!htp]
\centering
\includegraphics[width=\textwidth]{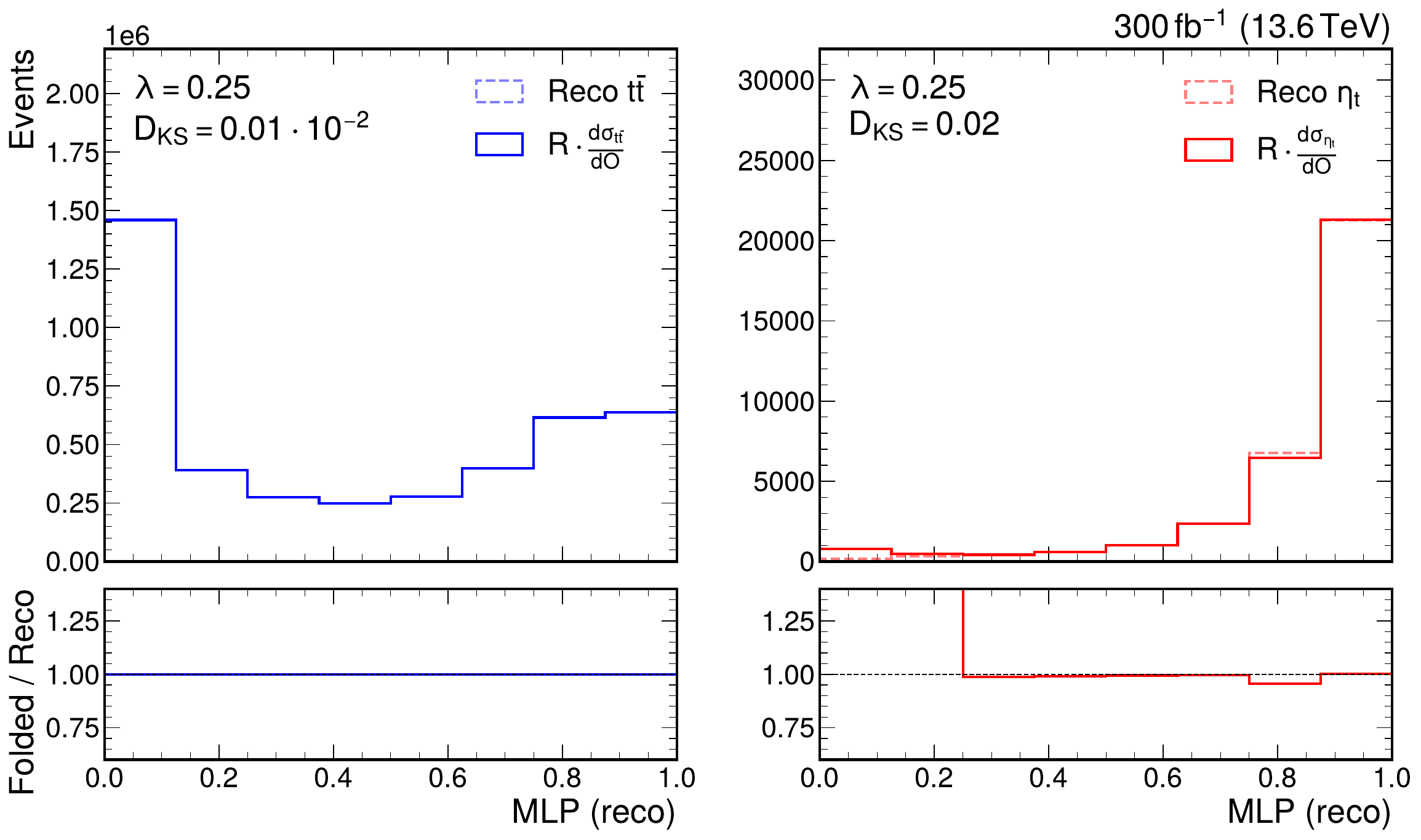}
\caption{%
    Comparison of the \ttbar (left, blue) and \etat (right, red) yields between \xO (dashed) and the folded distribution $R\,\dSdOinl$ (solid).
    The lower panel displays the ratio between the two distributions.
    Shown are the results for the training using the response-matrix constraint with $\lambda=0.25$.
    The captions list the Kolmogorov--Smirnov distance values.
}
\label{fig:biaslambda0p25}
\end{figure}

To reduce this bias, we include the response-matrix constraint from Eq.~\eqref{eq:responseconstraint} in the definition of the loss function.
The strength $\lambda$ of the constraint is then a hyperparameter of the training, and values $\lambda>0$ reduce the bias.
The comparison of the detector-level distributions for the example of a training with $\lambda=0.25$ is shown in Fig.~\ref{fig:biaslambda0p25}, showing a significantly reduced bias for the \etat production distribution.

\begin{figure}[!htp]
\centering
\includegraphics[width=0.475\textwidth]{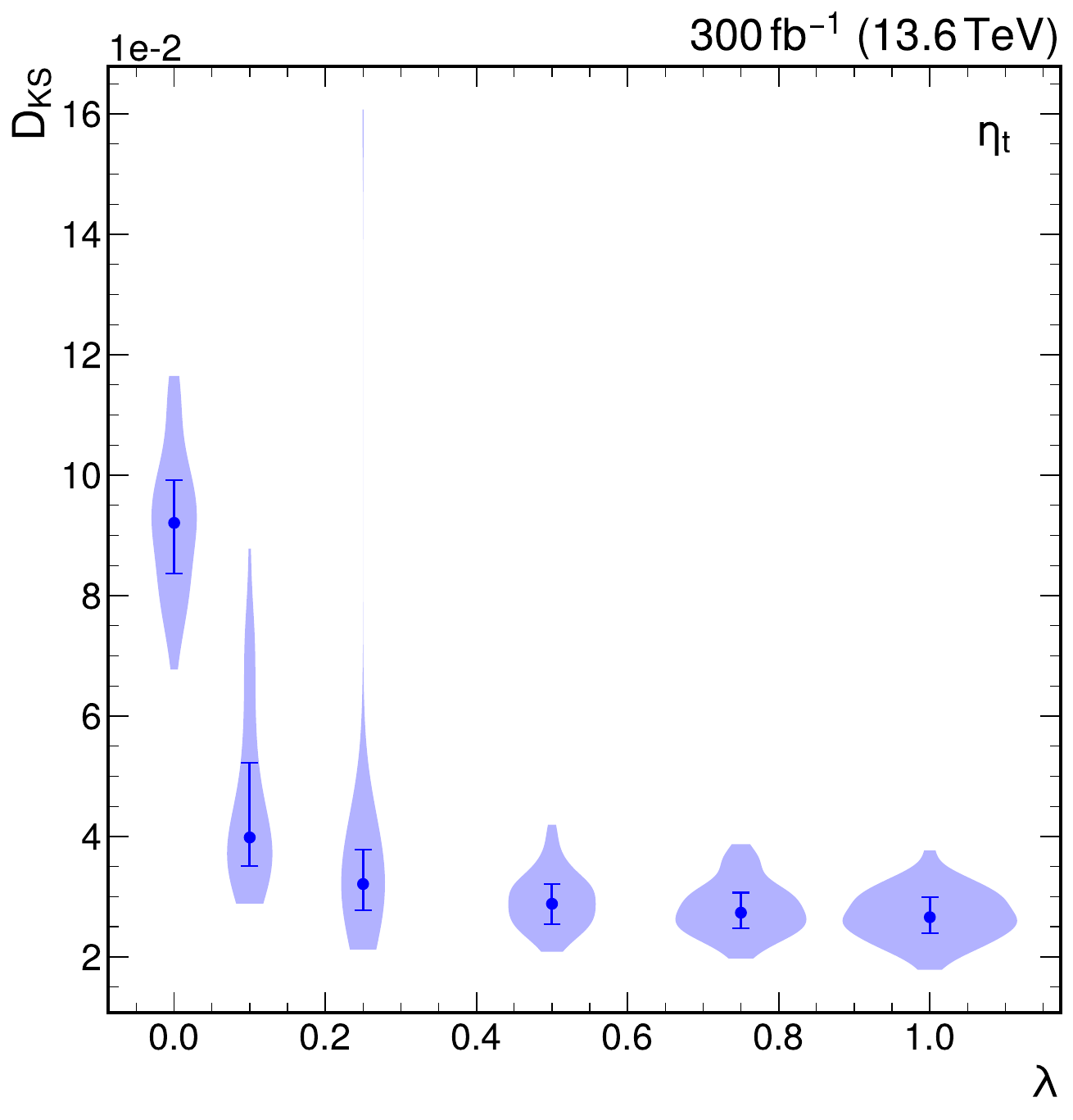}
\hfill
\includegraphics[width=0.49\textwidth]{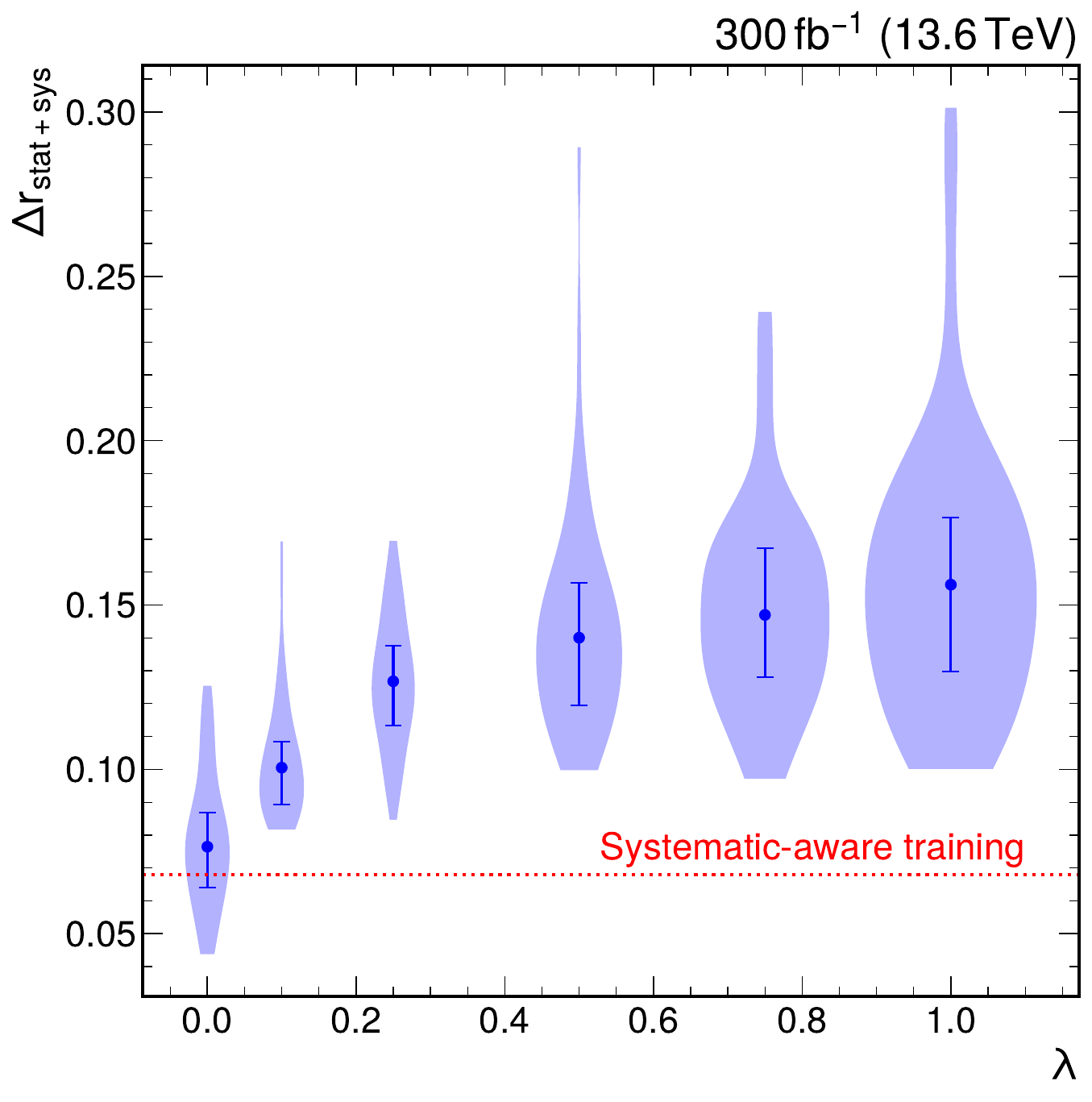}
\caption{%
    Results for different values of the response-matrix constraint strength $\lambda$. The violin plots show the distribution of the 50 trainings per $\lambda$ value, the dots the median value, and the error bars the 25 and 75\% quantiles.
    Left: Kolmogorov--Smirnov distance for the agreement between \xO and the folded distribution $R\,\dSdOinl$ shown for the \etat (right) yields.
    Right: Uncertainty in the \etat signal strength. The dotted red line indicates the uncertainty as obtained by the systematic-uncertainty-aware training shown in Fig.~\ref{fig:detectorlevelresults}.
}
\label{fig:ks_over_lambda}
\end{figure}

We perform the training systematically for a range of possible $\lambda$ values.
For each value of $\lambda$, the training is performed 50 times to average out the randomness due to the initialization of the MLPs.
To identify an optimal choice of $\lambda$, we evaluate both the bias via the value of $D_{\mathrm{KS}}$ and the overall precision in $r$ obtained by evaluating Eq.~\eqref{eq:loss} assuming the likelihood in Eq.~\eqref{eq:mainlikelihood}.
Figure~\ref{fig:ks_over_lambda} displays the $D_{\mathrm{KS}}$ results (left) and the expected precision (right).

For $\lambda=0$, the values of $\Delta r$ are comparable with the results obtained by the systematic-uncertainty-aware training indicated by the red dotted line.
While the introduction of the response-matrix constraint clearly reduces the bias and thus makes the unfolding less dependent on the parameter of interest $r$, this comes at the expense of sensitivity.
For this application, a value of $\lambda=0.25$ provides a good balance between high precision and low bias.

\section{Conclusions}

We present the \textit{Optimal Observable Machine} (OOM), a machine-learning-based framework for the construction of generator-level observables optimized for parameter extraction.
The key feature of the OOM is the explicit inclusion of detector response and systematic uncertainties in a likelihood-based training procedure, allowing the learned observable to be optimized directly for the expected measurement precision while remaining suitable for unfolding and reinterpretation.

The method combines detector-level and generator-level learnable functions and enforces consistency between them through the response matrix.
To control biases arising from a parameter-dependent response, we introduce an explicit constraint that suppresses such dependencies.
This enables the construction of observables whose sensitivity to the parameter of interest predominantly originates from the generator-level distribution, thereby preserving the conceptual advantages of unfolded measurements.

As a proof of principle, we apply the OOM to the measurement of a pseudoscalar excess at the top quark pair production threshold in dilepton final states.
We demonstrate that the method yields a generator-level observable with enhanced sensitivity compared to traditional approaches, while maintaining a controlled dependence of the unfolding on the signal strength.
The trade-off between bias and precision introduced by the response-matrix constraint was studied quantitatively, and an optimal working point was identified.

The OOM provides a general strategy for reconciling the high precision achievable with modern machine-learning techniques with the long-term interpretability of experimental results.
As a next step, symbolic regression~\cite{Butter:2021rvz, AbdusSalam:2024obf, Dong:2022trn} can be applied to the learned generator-level distribution to obtain an analytic formulation of the observable, thus facilitating reinterpretations of the unfolded results without need for the implementation of the machine-learning models.
While illustrated here in the context of top-quark physics, the method is broadly applicable to precision measurements and new-physics searches where unfolded observables are desired.

\acknowledgments

The authors thank the CERN summer student program, within which part of this work was carried out.
In particular, T.M.\ and F.R.\ acknowledge the opportunity to be summer students at CERN.
Ja.K.\ was supported by the Alexander-von-Humboldt foundation during the initial phase of the project.

\newcommand{\DOI}[1]{\href{http://dx.doi.org/#1}{\doi{#1}}}
\bibliography{refs.bib}

@article{AbdusSalam:2024obf,
    author = "AbdusSalam, S. and Abel, S. and Crispim Rom{\~a}o, M.",
    title = "Symbolic regression for beyond the standard model physics",
    eprint = "2405.18471",
    archivePrefix = "arXiv",
    primaryClass = "hep-ph",
    doi = "10.1103/PhysRevD.111.015022",
    journal = "Phys. Rev. D",
    volume = "111",
    pages = "015022",
    year = "2025"
}

@article{ALICE:2012cor,
    author = "Abelev, B. and others",
    collaboration = "ALICE",
    title = "Transverse sphericity of primary charged particles in minimum bias proton-proton collisions at $\sqrt{s}=0.9$, 2.76 and {7\TeV}",
    eprint = "1205.3963",
    archivePrefix = "arXiv",
    primaryClass = "hep-ex",
    doi = "10.1140/epjc/s10052-012-2124-9",
    journal = "Eur. Phys. J. C",
    volume = "72",
    pages = "2124",
    year = "2012"
}

@article{Alwall:2014hca,
    author = "Alwall, J. and Frederix, R. and Frixione, S. and Hirschi, V. and Maltoni, F. and Mattelaer, O. and Shao, H.-S. and Stelzer, T. and Torrielli, P. and Zaro, M.",
    title = "The automated computation of tree-level and next-to-leading order differential cross sections, and their matching to parton shower simulations",
    eprint = "1405.0301",
    archivePrefix = "arXiv",
    primaryClass = "hep-ph",
    doi = "10.1007/JHEP07(2014)079",
    journal = "JHEP",
    volume = "07",
    pages = "079",
    year = "2014"
}

@unpublished{Arcadi:2025grl,
    author = "Arcadi, G. and Djouadi, A.",
    title = "Interpreting the current {Higgs} excesses at the {LHC} in the {2HD}+{\Pa} framework",
    eprint = "2512.08807",
    archivePrefix = "arXiv",
    primaryClass = "hep-ph",
    year = "2025"
}

@article{Artoisenet:2012st,
    author = "Artoisenet, P. and Frederix, R. and Mattelaer, O. and Rietkerk, R.",
    title = "Automatic spin-entangled decays of heavy resonances in {Monte Carlo} simulations",
    eprint = "1212.3460",
    archivePrefix = "arXiv",
    primaryClass = "hep-ph",
    doi = "10.1007/JHEP03(2013)015",
    journal = "JHEP",
    volume = "03",
    pages = "015",
    year = "2013"
}

@article{ATLAS:2010arf,
    author = "Aad, G. and others",
    collaboration = "ATLAS",
    title = "The {ATLAS} simulation infrastructure",
    eprint = "1005.4568",
    archivePrefix = "arXiv",
    primaryClass = "physics.ins-det",
    doi = "10.1140/epjc/s10052-010-1429-9",
    journal = "Eur. Phys. J. C",
    volume = "70",
    pages = "823",
    year = "2010"
}

@techreport{ATLAS:2011tau,
    author = "{ATLAS and CMS Collaborations, and LHC Higgs Combination Group}",
    title = "Procedure for the {LHC} {Higgs} boson search combination in {Summer} 2011",
    number = "CMS-NOTE-2011-005, ATL-PHYS-PUB-2011-11",
    year = "2011",
    url = "https://cds.cern.ch/record/1379837"
}

@article{ATLAS:2020cli,
    author = "Aad, G. and others",
    collaboration = "ATLAS",
    title = "Jet energy scale and resolution measured in proton-proton collisions at $\sqrt{s}={13\TeV}$ with the {ATLAS} detector",
    eprint = "2007.02645",
    archivePrefix = "arXiv",
    primaryClass = "hep-ex",
    doi = "10.1140/epjc/s10052-021-09402-3",
    journal = "Eur. Phys. J. C",
    volume = "81",
    pages = "689",
    year = "2021"
}

@article{ATLAS:2024fdw,
    author = "Aad, G. and others",
    collaboration = "ATLAS",
    title = "Exploration at the high-energy frontier: {ATLAS} \mbox{Run 2} searches investigating the exotic jungle beyond the standard model",
    eprint = "2403.09292",
    archivePrefix = "arXiv",
    primaryClass = "hep-ex",
    doi = "10.1016/j.physrep.2024.10.001",
    journal = "Phys. Rept.",
    volume = "1116",
    pages = "301",
    year = "2025"
}

@article{ATLAS:2024fkg,
    author = "Aad, G. and others",
    collaboration = "ATLAS",
    title = "Characterising the {Higgs} boson with {ATLAS} data from the {LHC} \mbox{Run 2}",
    eprint = "2404.05498",
    archivePrefix = "arXiv",
    primaryClass = "hep-ex",
    doi = "10.1016/j.physrep.2024.11.001",
    journal = "Phys. Rept.",
    volume = "1116",
    pages = "4",
    year = "2025"
}

@article{ATLAS:2024itc,
    author = "Aad, G. and others",
    collaboration = "ATLAS",
    title = "{ATLAS} searches for additional scalars and exotic {Higgs} boson decays with the {LHC} \mbox{Run 2} dataset",
    eprint = "2405.04914",
    archivePrefix = "arXiv",
    primaryClass = "hep-ex",
    doi = "10.1016/j.physrep.2024.09.002",
    journal = "Phys. Rept.",
    volume = "1116",
    pages = "184",
    year = "2025"
}

@article{ATLAS:2024kxj,
    author = "Aad, G. and others",
    collaboration = "ATLAS",
    title = "Climbing to the top of the {ATLAS} {13\TeV} data",
    eprint = "2404.10674",
    archivePrefix = "arXiv",
    primaryClass = "hep-ex",
    doi = "10.1016/j.physrep.2024.12.004",
    journal = "Phys. Rept.",
    volume = "1116",
    pages = "127",
    year = "2025"
}

@article{ATLAS:2024lda,
    author = "Aad, G. and others",
    collaboration = "ATLAS",
    title = "The quest to discover supersymmetry at the {ATLAS} experiment",
    eprint = "2403.02455",
    archivePrefix = "arXiv",
    primaryClass = "hep-ex",
    doi = "10.1016/j.physrep.2024.09.010",
    journal = "Phys. Rept.",
    volume = "1116",
    pages = "261",
    year = "2025"
}

@article{ATLAS:2024wla,
    author = "Aad, G. and others",
    collaboration = "ATLAS",
    title = "Electroweak, {QCD} and flavour physics studies with {ATLAS} data from \mbox{Run 2} of the {LHC}",
    eprint = "2404.06829",
    archivePrefix = "arXiv",
    primaryClass = "hep-ex",
    doi = "10.1016/j.physrep.2024.12.003",
    journal = "Phys. Rept.",
    volume = "1116",
    pages = "57",
    year = "2025"
}

@techreport{ATLAS:2025mvr,
    author = "{ATLAS Collaboration}",
    collaboration = "ATLAS",
    title = "Observation of a cross-section enhancement near the \ttbar production threshold in $\sqrt{s}={13\TeV}$ ${\Pp\Pp}$ collisions with the {ATLAS} detector",
    type = "ATLAS Conference Note",
    number = "ATLAS-CONF-2025-008",
    year = "2025",
    url = "https://cds.cern.ch/record/2937636"
}

@article{Bahl:2025you,
    author = "Bahl, H. and Kumar, R. and Weiglein, G.",
    title = "Impact of interference effects on {Higgs}-boson searches in the di-top final state at the {LHC}",
    eprint = "2503.02705",
    archivePrefix = "arXiv",
    primaryClass = "hep-ph",
    doi = "10.1007/JHEP05(2025)098",
    journal = "JHEP",
    volume = "05",
    pages = "098",
    year = "2025"
}

@article{Bierlich:2022pfr,
    author = "Bierlich, C. and Chakraborty, S. and Desai, N. and Gellersen, L. and Helenius, I. and Ilten, P. and L{\"o}nnblad, L. and Mrenna, S. and Prestel, S. and Preuss, C. T. and Sj{\"o}strand, T. and Skands, P. and Utheim, M. and Verheyen, R.",
    title = "A comprehensive guide to the physics and usage of {\PYTHIA8.3}",
    eprint = "2203.11601",
    archivePrefix = "arXiv",
    primaryClass = "hep-ph",
    doi = "10.21468/SciPostPhysCodeb.8",
    journal = "SciPost Phys. Codeb.",
    volume = "8",
    year = "2022"
}

@article{Butter:2021rvz,
    author = "Butter, A. and Plehn, T. and Soybelman, N. and Brehmer, J.",
    title = "Back to the formula---{LHC} edition",
    eprint = "2109.10414",
    archivePrefix = "arXiv",
    primaryClass = "hep-ph",
    doi = "10.21468/SciPostPhys.16.1.037",
    journal = "SciPost Phys.",
    volume = "16",
    pages = "037",
    year = "2024"
}

@article{Celada:2024mcf,
    author = "Celada, E. and Giani, T. and ter Hoeve, J. and Mantani, L. and Rojo, J. and Rossia, A. N. and Thomas, M. O. A. and Vryonidou, E.",
    title = "Mapping the {SMEFT} at high-energy colliders: from {LEP} and the {(HL-)LHC} to the {FCC}-${\Pe\Pe}$",
    eprint = "2404.12809",
    archivePrefix = "arXiv",
    primaryClass = "hep-ph",
    doi = "10.1007/JHEP09(2024)091",
    journal = "JHEP",
    volume = "09",
    pages = "091",
    year = "2024"
}

@article{CMS:2015rld,
    author = "Khachatryan, V. and others",
    collaboration = "CMS",
    title = "Measurement of the differential cross section for top quark pair production in ${\Pp\Pp}$ collisions at $\sqrt{s} $ = {8\TeV}",
    eprint = "1505.04480",
    archivePrefix = "arXiv",
    primaryClass = "hep-ex",
    doi = "10.1140/epjc/s10052-015-3709-x",
    journal = "Eur. Phys. J. C",
    volume = "75",
    pages = "542",
    year = "2015"
}

@article{CMS:2016lmd,
    author = "Khachatryan, V. and others",
    collaboration = "CMS",
    title = "Jet energy scale and resolution in the {CMS} experiment in ${\Pp\Pp}$ collisions at {8\TeV}",
    eprint = "1607.03663",
    archivePrefix = "arXiv",
    primaryClass = "hep-ex",
    doi = "10.1088/1748-0221/12/02/P02014",
    journal = "JINST",
    volume = "12",
    pages = "P02014",
    year = "2017"
}

@article{CMS:2019esx,
    author = "Sirunyan, A. M. and others",
    collaboration = "CMS",
    title = "Measurement of \ttbar normalised multi-differential cross sections in ${\Pp\Pp}$ collisions at $\sqrt{s}={13\TeV}$, and simultaneous determination of the strong coupling strength, top quark pole mass, and parton distribution functions",
    eprint = "1904.05237",
    archivePrefix = "arXiv",
    primaryClass = "hep-ex",
    doi = "10.1140/epjc/s10052-020-7917-7",
    journal = "Eur. Phys. J. C",
    volume = "80",
    pages = "658",
    year = "2020"
}

@article{CMS:2024bni,
    author = "Hayrapetyan, A. and others",
    collaboration = "CMS",
    title = "Review of searches for vector-like quarks, vector-like leptons, and heavy neutral leptons in proton-proton collisions at $\sqrt{s}={13\TeV}$ at the {CMS} experiment",
    eprint = "2405.17605",
    archivePrefix = "arXiv",
    primaryClass = "hep-ex",
    doi = "10.1016/j.physrep.2024.09.012",
    journal = "Phys. Rept.",
    volume = "1115",
    pages = "570",
    year = "2025"
}

@article{CMS:2024gzs,
    author = "Hayrapetyan, A. and others",
    collaboration = "CMS",
    title = "Stairway to discovery: A report on the {CMS} programme of cross section measurements from millibarns to femtobarns",
    eprint = "2405.18661",
    archivePrefix = "arXiv",
    primaryClass = "hep-ex",
    doi = "10.1016/j.physrep.2024.11.005",
    journal = "Phys. Rept.",
    volume = "1115",
    pages = "3",
    year = "2025"
}

@article{CMS:2024irj,
    author = "Hayrapetyan, A. and others",
    collaboration = "CMS",
    title = "Review of top quark mass measurements in {CMS}",
    eprint = "2403.01313",
    archivePrefix = "arXiv",
    primaryClass = "hep-ex",
    doi = "10.1016/j.physrep.2024.12.002",
    journal = "Phys. Rept.",
    volume = "1115",
    pages = "116",
    year = "2025"
}

@article{CMS:2024krd,
    author = "Hayrapetyan, A. and others",
    collaboration = "CMS",
    title = "Overview of high-density {QCD} studies with the {CMS} experiment at the {LHC}",
    eprint = "2405.10785",
    archivePrefix = "arXiv",
    primaryClass = "nucl-ex",
    doi = "10.1016/j.physrep.2024.11.007",
    journal = "Phys. Rept.",
    volume = "1115",
    pages = "219",
    year = "2025"
}

@article{CMS:2024onh,
    author = "Hayrapetyan, A. and others",
    collaboration = "CMS",
    title = "The {CMS} statistical analysis and combination tool: \textsc{combine}",
    eprint = "2404.06614",
    archivePrefix = "arXiv",
    primaryClass = "physics.data-an",
    doi = "10.1007/s41781-024-00121-4",
    journal = "Comput. Softw. Big Sci.",
    volume = "8",
    pages = "19",
    year = "2024"
}

@article{CMS:2024phk,
    author = "Hayrapetyan, A. and others",
    collaboration = "CMS",
    title = "Searches for {Higgs} boson production through decays of heavy resonances",
    eprint = "2403.16926",
    archivePrefix = "arXiv",
    primaryClass = "hep-ex",
    doi = "10.1016/j.physrep.2024.09.004",
    journal = "Phys. Rept.",
    volume = "1115",
    pages = "368",
    year = "2025"
}

@article{CMS:2024zqs,
    author = "Hayrapetyan, A. and others",
    collaboration = "CMS",
    title = "Dark sector searches with the {CMS} experiment",
    eprint = "2405.13778",
    archivePrefix = "arXiv",
    primaryClass = "hep-ex",
    doi = "10.1016/j.physrep.2024.09.013",
    journal = "Phys. Rept.",
    volume = "1115",
    pages = "448",
    year = "2025"
}

@article{CMS:2025cwy,
    author = "Chekhovsky, V. and others",
    collaboration = "CMS",
    title = "Development of systematic uncertainty-aware neural network trainings for binned-likelihood analyses at the {LHC}",
    eprint = "2502.13047",
    archivePrefix = "arXiv",
    primaryClass = "hep-ex",
    doi = "10.1140/epjc/s10052-025-14713-w",
    journal = "Eur. Phys. J. C",
    volume = "85",
    pages = "1360",
    year = "2025"
}

@article{CMS:2025dzq,
    author = "Hayrapetyan, A. and others",
    collaboration = "CMS",
    title = "Search for heavy pseudoscalar and scalar bosons decaying to a top quark pair in proton-proton collisions at $\sqrt{s}={13\TeV}$",
    eprint = "2507.05119",
    archivePrefix = "arXiv",
    primaryClass = "hep-ex",
    doi = "10.1088/1361-6633/ae2207",
    journal = "Rep. Prog. Phys.",
    volume = "88",
    pages = "127801",
    year = "2025"
}

@article{CMS:2025kzt,
    author = "Hayrapetyan, A. and others",
    collaboration = "CMS",
    title = "Observation of a pseudoscalar excess at the top quark pair production threshold",
    eprint = "2503.22382",
    archivePrefix = "arXiv",
    primaryClass = "hep-ex",
    doi = "10.1088/1361-6633/adf7d3",
    journal = "Rep. Prog. Phys.",
    volume = "88",
    pages = "087801",
    year = "2025"
}

@techreport{CMS:NOTE-2022-008,
    author = "{CMS Offline Software and Computing Group}",
    title = "{CMS} \mbox{Phase 2} computing model: update document",
    type = "CMS Note",
    number = "CMS-NOTE-2022-008",
    year = "2022",
    url = "https://cds.cern.ch/record/2815292"
}

@inproceedings{Cowan:2002in,
    author = "Cowan, G.",
    title = "A survey of unfolding methods for particle physics",
    booktitle = "{Proc. Conference on Advanced Statistical Techniques in Particle Physics: Durham, UK, March 18--22, 2002}",
    year = "2002",
    note = "[Conf. Proc. C 0203181 (2002) 248]",
    url = "https://www.ippp.dur.ac.uk/Workshops/02/statistics/proceedings//cowan.pdf"
}

@article{Cowan:2010js,
    author = "Cowan, G. and Cranmer, K. and Gross, E. and Vitells, O.",
    title = "Asymptotic formulae for likelihood-based tests of new physics",
    eprint = "1007.1727",
    archivePrefix = "arXiv",
    primaryClass = "physics.data-an",
    doi = "10.1140/epjc/s10052-011-1554-0",
    journal = "Eur. Phys. J. C",
    volume = "71",
    pages = "1554",
    year = "2011",
    note = "[Erratum: \DOI{10.1140/epjc/s10052-013-2501-z}]"
}

@unpublished{Cranmer:2015bka,
    author = "Cranmer, K. and Pavez, J. and Louppe, G.",
    title = "Approximating likelihood ratios with calibrated discriminative classifiers",
    eprint = "1506.02169",
    archivePrefix = "arXiv",
    primaryClass = "stat.AP",
    year = "2015"
}

@article{Czakon:2011xx,
    author = "Czakon, M. and Mitov, A.",
    title = "\textsc{top++}: a program for the calculation of the top-pair cross-section at hadron colliders",
    eprint = "1112.5675",
    archivePrefix = "arXiv",
    primaryClass = "hep-ph",
    doi = "10.1016/j.cpc.2014.06.021",
    journal = "Comput. Phys. Commun.",
    volume = "185",
    pages = "2930",
    year = "2014"
}

@unpublished{deBlas:2025xhe,
    author = "de Blas, J. and Goncalves, A. and Miralles, V. and Reina, L. and Silvestrini, L. and Valli, M.",
    title = "Constraining new physics effective interactions via a global fit of electroweak, {Drell}--{Yan}, {Higgs}, top, and flavour observables",
    eprint = "2507.06191",
    archivePrefix = "arXiv",
    primaryClass = "hep-ph",
    year = "2025"
}

@article{DeCastro:2018psv,
    author = "De Castro, P. and Dorigo, T.",
    title = "\textsc{inferno}: Inference-aware neural optimisation",
    eprint = "1806.04743",
    archivePrefix = "arXiv",
    primaryClass = "stat.ML",
    doi = "10.1016/j.cpc.2019.06.007",
    journal = "Comput. Phys. Commun.",
    volume = "244",
    pages = "170",
    year = "2019"
}

@article{deFavereau:2013fsa,
    author = "de Favereau, J. and Delaere, C. and Demin, P. and Giammanco, A. and Lema{\^\i}tre, V. and Mertens, A. and Selvaggi, M.",
    collaboration = "DELPHES 3",
    title = "\textsc{delphes} 3: a modular framework for fast simulation of a generic collider experiment",
    eprint = "1307.6346",
    archivePrefix = "arXiv",
    primaryClass = "hep-ex",
    doi = "10.1007/JHEP02(2014)057",
    journal = "JHEP",
    volume = "02",
    pages = "057",
    year = "2014"
}

@article{Dong:2022trn,
    author = "Dong, Z. and Kong, K. and Matchev, K. T. and Matcheva, K.",
    title = "Is the machine smarter than the theorist: Deriving formulas for particle kinematics with symbolic regression",
    eprint = "2211.08420",
    archivePrefix = "arXiv",
    primaryClass = "hep-ph",
    doi = "10.1103/PhysRevD.107.055018",
    journal = "Phys. Rev. D",
    volume = "107",
    pages = "055018",
    year = "2023"
}

@article{Fadin:1990wx,
    author = "Fadin, V. S. and Khoze, V. A. and Sj{\"o}strand, T.",
    title = "On the threshold behaviour of heavy top production",
    doi = "10.1007/BF01614696",
    journal = "Z. Phys. C",
    volume = "48",
    pages = "613",
    year = "1990"
}

@article{FerreiradaSilva:2023mhf,
    author = "Ferreira da Silva, P.",
    title = "Physics of the top quark at the {LHC}: An appraisal and outlook of the road ahead",
    doi = "10.1146/annurev-nucl-102419-052854",
    journal = "Ann. Rev. Nucl. Part. Sci.",
    volume = "73",
    pages = "255",
    year = "2023"
}

@article{Fisher:1922saa,
    author = "Fisher, R. A.",
    title = "On the mathematical foundations of theoretical statistics",
    doi = "10.1098/rsta.1922.0009",
    journal = "Phil. Trans. Roy. Soc. Lond. A",
    volume = "222",
    pages = "309",
    year = "1922"
}

@unpublished{Flacke:2025dwk,
    author = "Flacke, T. and Fuks, B. and Kim, D. and Kim, J. and Lee, S. J. and Munoz-Aillaud, L.",
    title = "New physics in toponium's shadow?",
    eprint = "2512.03220",
    archivePrefix = "arXiv",
    primaryClass = "hep-ph",
    year = "2025"
}

@article{Fuks:2021xje,
    author = "Fuks, B. and Hagiwara, K. and Ma, K. and Zheng, Y.-J.",
    title = "Signatures of toponium formation in {LHC} \mbox{Run 2} data",
    eprint = "2102.11281",
    archivePrefix = "arXiv",
    primaryClass = "hep-ph",
    doi = "10.1103/PhysRevD.104.034023",
    journal = "Phys. Rev. D",
    volume = "104",
    pages = "034023",
    year = "2021"
}

@article{Fuks:2024yjj,
    author = "Fuks, B. and Hagiwara, K. and Ma, K. and Zheng, Y.-J.",
    title = "Simulating toponium formation signals at the {LHC}",
    eprint = "2411.18962",
    archivePrefix = "arXiv",
    primaryClass = "hep-ph",
    doi = "10.1140/epjc/s10052-025-13853-3",
    journal = "Eur. Phys. J. C",
    volume = "85",
    pages = "157",
    year = "2025"
}

@unpublished{Fuks:2025toq,
    author = "Fuks, B. and Hossain, A. and Keaveney, J.",
    title = "Statistical indications of toponium formation in top quark pair production",
    eprint = "2511.02040",
    archivePrefix = "arXiv",
    primaryClass = "hep-ph",
    year = "2025"
}

@article{Garzelli:2024uhe,
    author = "Garzelli, M. V. and Limatola, G. and Moch, S.-O. and Steinhauser, M. and Zenaiev, O.",
    title = "Updated predictions for toponium production at the {LHC}",
    eprint = "2412.16685",
    archivePrefix = "arXiv",
    primaryClass = "hep-ph",
    doi = "10.1016/j.physletb.2025.139532",
    journal = "Phys. Lett. B",
    volume = "866",
    pages = "139532",
    year = "2025"
}

@article{GEANT4:2002zbu,
    author = "Agostinelli, S. and others",
    collaboration = "GEANT4",
    title = "{\GEANTfour}---a simulation toolkit",
    doi = "10.1016/S0168-9002(03)01368-8",
    journal = "Nucl. Instrum. Meth. A",
    volume = "506",
    pages = "250",
    year = "2003"
}

@article{Good:1952rss,
    author = "Good, I. J.",
    title = "Rational decisions",
    doi = "10.1111/j.2517-6161.1952.tb00104.x",
    journal = "J. Royal Stat. Soc. B",
    volume = "14",
    pages = "107",
    year= "1952"
}

@article{Maltoni:2024tul,
    author = "Maltoni, F. and Severi, C. and Tentori, S. and Vryonidou, E.",
    title = "Quantum detection of new physics in top-quark pair production at the {LHC}",
    eprint = "2401.08751",
    archivePrefix = "arXiv",
    primaryClass = "hep-ph",
    doi = "10.1007/JHEP03(2024)099",
    journal = "JHEP",
    volume = "03",
    pages = "099",
    year = "2024"
}

@unpublished{Matsuoka:2025jgm,
    author = "Matsuoka, Y.",
    title = "Possible mixing between elementary and bound state fields in the \ttbar production excess at the {LHC}",
    eprint = "2510.16828",
    archivePrefix = "arXiv",
    primaryClass = "hep-ph",
    year = "2025"
}

@inproceedings{Mertens:2015kba,
    author = "Mertens, A.",
    title = "New features in \textsc{delphes} 3",
    booktitle = "{Proc. 16th International Workshop on Advanced Computing and Analysis Techniques in Physics Research (ACAT 2014): Prague, Czechia, September 1--5, 2014}",
    doi = "10.1088/1742-6596/608/1/012045",
    year = "2015",
    note = "[J. Phys. Conf. Ser. 608 (2015) 012045]"
}

@inproceedings{Nair:2010icml,
    author = "Nair, V. and Hinton, G. E.",
    title="Rectified linear units improve restricted {Boltzmann} machines",
    booktitle = "{Proc. 27th International Conference on IMachine Learning (ICML 2010): Haifa, Israel, June 21--24, 2010}",
    pages = "807",
    year = "2010",
    url = "https://www.cs.toronto.edu/~fritz/absps/reluICML.pdf"
}

@article{Nason:2025hix,
    author = "Nason, P. and Re, E. and Rottoli, L.",
    title = "Spin correlations in \ttbar production and decay at the {LHC} in {QCD} perturbation theory",
    eprint = "2505.00096",
    archivePrefix = "arXiv",
    primaryClass = "hep-ph",
    doi = "10.1007/JHEP10(2025)149",
    journal = "JHEP",
    volume = "10",
    pages = "149",
    year = "2025"
}

@inproceedings{Neal:2007zz,
    author = "Neal, R. M.",
    title = "Computing likelihood functions for high-energy physics experiments when distributions are defined by simulators with nuisance parameters",
    booktitle = "{Proc. LHC Workshop on Statistical Issues for LHC Physics (PHYSTAT-LHC): Geneva, Switzerland, June 27--29, 2007}",
    doi = "10.5170/CERN-2008-001.111",
    pages = "111",
    year = "2007",
}

@article{NNPDF:2017mvq,
    author = "Ball, R. D. and Bertone, V. and Carrazza, S. and Del Debbio, L. and Forte, S. and Groth-Merrild, P. and Guffanti, A. and Hartland, N. P. and Kassabov, Z. and Latorre, J. I. and Nocera, E. R. and Rojo, J. and Rottoli, L. and Slade, E. and Ubiali, M.",
    collaboration = "NNPDF",
    title = "Parton distributions from high-precision collider data",
    eprint = "1706.00428",
    archivePrefix = "arXiv",
    primaryClass = "hep-ph",
    doi = "10.1140/epjc/s10052-017-5199-5",
    journal = "Eur. Phys. J. C",
    volume = "77",
    pages = "663",
    year = "2017"
}

@inproceedings{Paszke:2019xhz,
    author = "Paszke, A. and others",
    title = "\textsc{PyTorch}: An imperative style, high-performance deep learning library",
    booktitle = "{Proc. 33rd Conference on Neural Information Processing Systems (NeurIPS 2019): Vancouver, Canada, December 08--14, 2019}",
    eprint = "1912.01703",
    archivePrefix = "arXiv",
    primaryClass = "cs.LG",
    year = "2019"
}

@article{Schwartz:2021ftp,
    author = "Schwartz, M. D.",
    title = "Modern machine learning and particle physics",
    eprint = "2103.12226",
    archivePrefix = "arXiv",
    primaryClass = "hep-ph",
    doi = "10.1162/99608f92.beeb1183",
    journal = "Harv. Data Sci. Rev.",
    volume = "3",
    pages = "2",
    year = "2021"
}

@inproceedings{Selvaggi:2014mya,
    author = "Selvaggi, M.",
    title = "\textsc{delphes} 3: A modular framework for fast-simulation of generic collider experiments",
    booktitle = "{Proc. 15th International Workshop on Advanced Computing and Analysis Techniques in Physics Research (ACAT 2013): Beijing, China, May 16--21, 2013}",
    doi = "10.1088/1742-6596/523/1/012033",
    year = "2014",
    note = "[J. Phys. Conf. Ser. 523 (2014) 012033]"
}

@inproceedings{Simpson:2022suz,
    author = "Simpson, N. and Heinrich, L.",
    title = "\textsc{neos}: End-to-end-optimised summary statistics for high energy physics",
    booktitle = "{Proc. 20th International Workshop on Advanced Computing and Analysis Techniques in Physics Research (ACAT 2021): Daejeon, South Korea, November 29--December 03, 2021}",
    eprint = "2203.05570",
    archivePrefix = "arXiv",
    primaryClass = "physics.data-an",
    doi = "10.1088/1742-6596/2438/1/012105",
    year = "2023",
    note = "[J. Phys. Conf. Ser. 2438 (2023) 012105]"
}

@article{Sumino:2010bv,
    author = "Sumino, Y. and Yokoya, H.",
    title = "Bound-state effects on kinematical distributions of top quarks at hadron colliders",
    eprint = "1007.0075",
    archivePrefix = "arXiv",
    primaryClass = "hep-ph",
    doi = "10.1007/JHEP09(2010)034",
    journal = "JHEP",
    volume = "09",
    pages = "034",
    year = "2010",
    note = "[Erratum: \DOI{10.1007/JHEP06(2016)037}]"
}

@article{Wunsch:2018oxb,
    author = "Wunsch, S. and Friese, R. and Wolf, R. and Quast, G.",
    title = "Identifying the relevant dependencies of the neural network response on characteristics of the input space",
    eprint = "1803.08782",
    archivePrefix = "arXiv",
    primaryClass = "physics.data-an",
    doi = "10.1007/s41781-018-0012-1",
    journal = "Comput. Softw. Big Sci.",
    volume = "2",
    pages = "5",
    year = "2018"
}

@article{Wunsch:2020iuh,
    author = "Wunsch, S. and J{\"o}rger, S. and Wolf, R. and Quast, G.",
    title = "Optimal statistical inference in the presence of systematic uncertainties using neural network optimization based on binned {Poisson} likelihoods with nuisance parameters",
    eprint = "2003.07186",
    archivePrefix = "arXiv",
    primaryClass = "physics.data-an",
    doi = "10.1007/s41781-020-00049-5",
    journal = "Comput. Softw. Big Sci.",
    volume = "5",
    pages = "4",
    year = "2021"
}

\end{document}